# EconoThermodynamics, or the world economy "thermal death" paradox.

**A.M. Tishin,**

Physics Department of M.V. Lomonosov Moscow State University, Leninskie Gory, Moscow 119992

**O.B. Baklitskaya*,**

Polymagnet LLC, AMT&C Group, Moscow, Russia

**Abstract**

The paper present one of attempts to apply the thermodynamics laws to economics. Introducing common thermodynamic parameters and considering world economics as a one macrosystem, authors demonstrate the possible consequences of entropy increasing due to irreversible economics activities. "Entropy" advices to leaders of different business units are presented.

**Preface**

*"Yes, the world is changing and it's going to continue to change. But that does not mean we should stop the search for timeless principles. Think of it this way: while the practices of engineering continually evolve and change, the laws of physics remain relatively fixed. I like to think of our work as a search for timeless principles, the enduring physics of great organizations that will remain true and relevant no matter how the world changes around us. Yes, the specific application will change, the engineering, if you will, but certain immutable laws of organized human performance - the physics - will endure"[1].*

**Introduction. ECONOPHYSICS is 10 years old**

Any physicist working in business, economics and finance will agree that his physics education has helped him to understand the laws governing these areas. Is this just what one would expect or is there something more to it than that? Physics and economics, thermodynamics and business - what do they have in common? At first sight, nothing at all. The rapid development of different types of technology, permeating everyday life at an ever increasing rate, demands vast scientific knowledge from business people. The enormous amount of information that has became available to researchers, thanks to the development of internet technology, has led to "cross-pollination" between different sciences - economics, biology, mathematics, ecology, physics and computer technology. There is one obvious example: over the last 10 years, the natural sciences have become so closely intertwined with economics that they have given birth to a new interdisciplinary field named **econophysics**. Its official date of birth is given as 1997, when the first econophysics conference was held in Budapest.

---

* Corresponding author: baklitskaya@amtc.ru

If we look at mathematics, we can see that its permeation of economics over the course of the last 150 years has been highly successful. As early as the end of the 19th century, the Italian engineer-economist, Vilfredo Pareto, discovered a universal law which was true not only for both physics and chemistry, but also for the distribution of material wealth [2]. He deduced that the composite probability of high income in Western societies follows the inverse power law. (This relationship is confirmed by statistical graphs for the distribution of material wealth in most countries).

$$P(w) = \int_w^\infty p(x)\,dx \propto \frac{1}{w^\mu},$$

where *P(w)* is the probability of finding an individual, within a given economic system, with an income higher than value *w,* defining material wealth, and *p(x)* is the density function of such an occurrence. The value *w* (prosperity, material wealth) is the aggregate value of the individual's property (intellectual property, real estate, investment income, etc.), and $1 \leq \mu \leq 2$.

Pareto's law inspired scientists to search for a mathematical way of predicting economic occurrences, and controlling them. One example of this is the Club of Rome report, presented by a group led by Prof. D. Meadows [3], on using computer forecasting to forecast world development. Mathematical theories and models are successfully applied to economics today, examples being mathematical statistics, optimal control theory, convex analysis, topology and stability theory, etc. To solve actual business problems, however, it is impossible to do without a synthesis of the laws and principles of economathematics with those of physics.

The enormous experience gained by physicists from analyzing experimental research data does in fact give them a distinct advantage when it comes to uncovering numerical laws governing economic and financial statistics. The interdisciplinary field of econophysics is opening up new possibilities, capable of revolutionizing economic and social science disciplines. The followers of the new discipline (and these are mainly physicists, not economists) are creating new methods in complex system physics and statistical analysis in order to describe a wide array of empirical data gathered in the economics field.

As an example, an analysis, produced by Prof. V.Yakovenko [4], of the data of the US Internal Revenue Service from 1983-2001 showed that methods used in statistical physics can be successfully applied to solve economic and financial problems. In particular, the team widened the scope of Pareto's studies of rich and poor people, and demonstrated that the wealth of super-rich people (about 3%) follows the Pareto law (the number of wealthy people is **W** ~1/**W**$^e$, where **e** is in the range of 2 to 3), while the wealth of the remaining 97% corresponds exactly to the distribution of energy in gas atoms [5].

The number of conferences on econophysics and the application of physics methods to economics increases with each passing year. Whereas the first conference in Budapest in 1997 discussed only two theoretical questions - could physics contribute to economic and financial modeling, and could it have something to say about market fluctuation - nowadays econophysics is getting more and more involved in practical issues. The sixth International Conference, entitled The Application of Physics to Financial Analysis, was held in Lisbon in July 2007, and attracted more than 100 participants, including Russian scientists. Among the topics discussed were problems of financial markets and how to model them, risk control, games theory, macroeconomic models, scaling and equilibrium processes, network theory, and methods of quantum statistics in financial questions.

There is currently an avalanche of studies of the analysis and prognosis of various economic processes. This is largely due to the increased processing power of the computers used for economic modeling. However, despite the optimistic predictions for using automated control systems, reality has fallen short of expectation. Many numerical models do not work adequately because of insufficient statistical data and because there are not enough equations to describe their behavior. Furthermore, it can be hard to define the parameters of a problem clearly when they may change radically following a single announcement from a politician.

Of course, financial experts are attracted by the possibility of applying advanced control models and probabilistic forecasting to economic situations, but frequently there is insufficient mutual understanding between economists and physicists. Nevertheless, productive cooperation is possible. Special econophysics courses or combined courses of physics and economics are currently available universities.

Amongst the successful collaborations between the natural sciences (especially mathematics) and economics, there is one which clearly stands out, and this is the synthesis of economics with another fundamental science - **thermodynamics**.

The differences between physics and other sciences are tentative and somewhat diffuse. That is why many universities now do not have separate buildings for departments of physics, chemistry and other sciences, and often physicists, mathematicians, chemists, biologists, medics and other scientists simply collaborate on the same projects. Working in the field of nanotechnology, for example, they use the same atomic force, electron and scanning microscopes (possibly with different optional equipment).

But does economics share a common "border" with physics, and is there therefore a shared "border area" between the two? Or does it only "border on" such natural sciences as mathematics and statistics? But physics and economics are, after all, equally reliant on mathematics and statistics. In our view, there is no doubt that a shared "border" exists and, consequently, that there is also a "border area" for shared scientific research in physics and economics. That is why physicists give science lectures in the Department of Economics of Moscow State University. Unlike economics, where experiments are extremely difficult and dangerous to carry out, physics can hardly be imagined without them. The laws of physics are based on facts acquired experimentally, even in cases where a full understanding of theory has not been reached - for example, in high-temperature superconductivity. Is it possible to use some methods and approaches from physics to describe economic processes? We believe that such a possibility could contribute to a "physics" understanding of these processes, based on experimental practice.

But which methods from physics can be used to describe economic processes? Econophysics attempts to find the answer to this question. In the present article, we review the possibility of applying a thermodynamic approach to the description of macroeconomics, as this is the approach most generally applicable to the characterization of the behavior of physical systems consisting of a large population of objects.

Why did we choose to begin with thermodynamics and not quantum physics? As is well known, thermodynamics is not interested in the nature of interactions and processes on a micro-level, so we do not have to answer the question we are likely to be asked: "Do you know how the individual elements of macroeconomic systems interact with each other?" Econothermodynamics is not concerned with that. Remember that ever since the appearance of the first work by its founder, Sadi Carnot, in 1824, thermodynamics has been defined as the science of "energy in transit" as seen during thermal processes.

We are currently living through a slowdown in the growth of the world economy, an era of crises for different economies, including those of developed countries. Leading economists today admit that no-one has yet managed to create a successful economic theory, capable of describing adequately the macroeconomic processes currently at work.

Let us return to Pareto's "prosperity" formula. It implies that any society comprises more poor people than wealthy. As noted by K. Bogdanov [6], the relative fraction of people ΔF, possessing wealth between **D – ΔD/2** and **D + ΔD/2**, can be calculated using this formula:

$$\Delta F = \frac{e^{-\frac{D}{D_a}}}{D_a} \cdot \Delta D, \qquad (1)$$

where $D_a$ is the average annual income of U.S. citizens, equal to 23.2 thousand USD (1996).

The formula (1) for incomes in a society is similar to the well-known Boltzmann-Gibbs-Maxwell distribution equation from thermodynamics (2). It defines the relative fraction ΔF of gas molecules having temperature of **T**, whose potential energy is within the range of **E±dE/2**:

$$\Delta F = \frac{e^{-\frac{E}{kT}}}{kT} \cdot \Delta E. \quad (2)$$

These two equations, from different fields, have one thing in common: they follow the conservation laws. When gas molecules collide, their total mechanical energy is transferred from one molecule to the other without undergoing any change. The same law applies to the interaction between the seller and the buyer of goods or services. After the deal is done, the wealth of one of them (the buyer or the seller) increases, while the wealth of the other decreases by the same amount. One of the reasons this cannot be called "robbery" is that the price of goods or services is not always easy to determine.

## ECONOTHERMODYNAMICS

The concept of biosphere entropy came into existence long ago. Back in 1964, K.S. Trincher formulated the theorem of its growth threshold (see for example [7]). But today, the Earth's biosphere has actually become part of its economic processes, being no longer simply its "lungs", but also one of its types of "treatment facility". To what extent are the laws of thermodynamics true not only of the biosphere, but of economics as a whole? Economics can be regarded as a system (for the moment it does not matter whether a closed or an open one), which cannot be described without an understanding of the entropy it produces (the system irregularity factor). And according to the second law of thermodynamics, all real physical processes are irreversible. A formal characterization of economics based on thermodynamics was suggested several years ago by a J. Mimkes [8], using what are essentially statistical methods. In his work, the laws of thermodynamics are formulated, not in conventional physical terms (heat, work, internal energy), but rather in terms accepted in economic and financial circles: capital growth, labor costs, value added, etc. Applying the laws thus formulated to the German car market, car production and several other processes, the author shows that the ratios produced agree perfectly satisfactorily with the actual situation in these businesses. In the present work, in contrast those already mentioned, we will attempt to apply the thermodynamic approach to gain insight into the world economy on the macroeconomic level, as well as to review the specific behavior of particular business companies. Of course, we realize that the very possibility of applying such an approach demands further substantiation, and that the conclusions reached demand verification based on experimental facts. Also, it should be emphasized that the authors of this work are physicists, not economists - they are well acquainted with laws of physics, but have no in-depth knowledge of economics.

The question of whether an economy is an open or a closed thermodynamic system demands further serious discussion and analysis. The economy of a particular country, trading "energy" and matter (goods), can certainly be seen as an open system. We have in mind here not only oil and gas, which, incidentally, can be seen as both energy and matter at the same time. It is worth noting that many of the world's recent wars and conflicts have been linked precisely to attempts to redistribute energy resources (oil and gas) and matter (minerals, precious metals, etc), and not only to ideological or religious causes. In the same way, energy and heat in thermodynamic systems are constantly changing and moving from one structural unit to another (for example, from a crystalline sublattice to a magnetic one, and from one part of heated body to another). Author of work [7] use the term of "entropy" to describe the biosphere as an open system. To counteract the constant growth of the biosphere's entropy, he proposes using solar energy as a source of "negative entropy". This could first be used to stop the growth of biosphere entropy, and then to attempt to return the biosphere to a state where its entropy would be constantly diminishing. In contrast to the biosphere, it is most probable that the world economy, as a whole, represents an almost closed system from a thermodynamic point of view. Although it does currently, to some extent, "consume" the energy of wind and sun, it does not, so far, exchange goods with, for example, Mars. Even a significant increase in the share of solar, wind and other types of renewable energy in the world's energy consumption (which will take at least 50 years), though it may improve the biosphere, will not mean a stop to the irreversible industrial and other activity of humans and machines. As we can see from recent events, price surges on the stock markets of Asia, the United States and Russia are closely linked, and the slow attenuation aftershock oscillations can last for a months scale time period, a feature characteristic of strongly correlated closed systems (when the locally arise in the system additional energy and/or substance have not possibility to leave out the close system and quickly normalize the situation).

Economics, like classic thermodynamics, is described on two levels. On one hand, there is micro-economics - the level at which the interaction between separate business entities takes place. Most previous works of econophysics are devoted to this level of description. On the other hand, when we talk about the behavior of the whole economic system, working as a complex entity, we are concerned with the macro-level. This level can be described using thermodynamic variables reflecting the activity of the system and the exchange of energy and matter between the system and the environment. In a conventional thermodynamic system, transformation of energy is caused by actual interactions and their regulation on the micro-level. The economic system is similar: the interaction between its separated micro-level parts leads to the formation of new processes on the macro-level. That is why the description of economic systems in thermodynamic terms can have a universal character. Naturally, just as in the case of conventional physical phenomena, econo-thermodynamics does not claim to describe micro-level processes. Its validity will have to be justified, in future, by a vast number of facts. We believe that quantum physics, condensed matter physics and other branches of physics which address patterns of interaction between separated macro-system objects also lend themselves, to some extent, to the description of micro-level processes.

There are several formulations of the second principle of thermodynamics. One of them is based on entropy (measure of the unavailability of a system's energy to do work) and states that, in an irreversible process, the entropy of an isolated system is not reduced (in other words, it either increases or remains constant upon reaching the maximum level). Could entropy, as a measure of system disorder, be used to describe the state of a macro-economy? We believe it can. It can also be used to describe the efficiency coefficient, familiar to everyone from his school days. As we know from thermodynamics, Carnot's theorems state that the efficiency coefficient of the Carnot cycle does not depend on the nature of the working matter or the terminal adiabats, but only on the temperatures of the hot and cold temperature reservoirs. The efficiency coefficient of any thermal engine does not exceed that of the Carnot cycle working under the same conditions. In other words, the efficiency coefficient of an irreversible engine is lower than that of a reversible engine. The goal of econothermodynamics is therefore to solve the main problem: how to increase the value of the efficiency coefficient of the macroeconomic process itself and bring it closer to 1. It should be noted, however, that the efficiency coefficient in this case will not depend on the actual macroeconomic process, but only on the temperatures of the hot and cold temperature reservoirs. Due to the fact that all the processes in the world economy or the economy of an individual country are irreversible, the efficiency coefficient of the world economy is not likely to reach 1, regardless of the nature of the process. One method of increasing the efficiency coefficient, familiar from physics, is to increase the temperature of the hot temperature reservoir; this method was successfully used by China to increase the efficiency coefficient of its economy. Since all macroeconomic processes are irreversible, it was impossible to establish communism (precisely the social structure which, in the view of Marxist-Leninist writers, is most likely to achieve an efficiency coefficient equal or close to 1) even with such an "overheated" economy as China's. History shows that a free market economy is much more effective - in other words, it has a higher efficiency coefficient - than a socialist one, and that it is unlikely that Marx's "Capital" can be used as a textbook for building a thermal engine with an efficiency coefficient approaching 1. However, in accord to Engels, the laws of conservation and transformation of energy are not applicable to the higher forms of the movement of matter.

When making the transition from classic thermodynamics to a thermodynamic characterization of the evolution and development of macroeconomic systems (econothermodynamics) it is necessary (depending on the macro-level of the system being studied) to review unbalanced, non-linear self-organizing systems with varying entropy (for example, quickly developing economies and countries). If a system self-organizes, its entropy reduces. In disorganizing, degrading systems, entropy increases.

Interestingly, Leo Szilard (Einstein's colleague) demonstrated back in the 1930s that "fluctuation processes can be described in terms of generalized, phenomenological, "continual" thermodynamics, without accounting for the atomic and molecular properties of matter". Szilard studied the conditions under which a rational being is able to break the second principle of thermodynamics, which states that it is impossible to build a cyclic thermal engine which can fully transform thermal energy into mechanical work. Furthermore, Szilard was the first to associate physics with informatics by characterizing the interaction between the latter and entropy (long before Wiener and Shannon) [9].

If the activity of the selected macro-system is based on natural principles, it will agree with the theorem of the Nobel prize winner, Ilya Prigozhin [10], to the effect that a system must tend to the state with a minimal increase of entropy.

The constant changes taking place during the vital activity of an economic macro-system cannot be interpreted as a simple transition from one balanced state to another. An economic macro-system, like a conventional thermodynamic system, maintains a state of thermodynamic balance only if the macroscopic values defining its state remain constant. The main goal of econothermodynamics is the precise definition of this set of econothermodynamic values (defining the state of the macro-system at any particular moment) and the interpretation of their economic meaning, and also finding out how the values for ideal and actual macroeconomic systems relate to one another. In conventional thermodynamics, the values in question are temperature, volume, pressure, and magnetic or electrical field. Below we suggest author's definitions for econothermodynamic analogues, such as internal and/or free energy, and the values they depend on.

So far, no suitable "thermodynamic" instrument has been developed which would allow the development of macroeconomic systems to be characterized correctly. This can be done in terms of the general theory of the evolution of non-linear complex systems, which evolve randomly and change into something different. A multiplicity of possible directions for development and the presence of so-called attractors (sets attracting different development directions) are characteristic for such systems. The possible directions for development are chosen by the system at bifurcation points, close to which the system is in an unbalanced, unstable state. At these moments, even a slight influence can cause significant changes in the system's behavior. Such systems achieve a balanced state only at isolated moments. In the unbalanced state, any non-linear system is susceptible to so-called resonance influences connected with its internal properties (later we will discuss this further).

While studying the states of an abstract (non-economic) developing system in terms of thermodynamics, Prigozhin came to the conclusion that reversible and irreversible processes produce different structures: balanced and unbalanced. It is the unbalanced ("dissipative") structures that are long-lived, due to the influx of energy and matter in an open system, so long as they are in a state with minimal entropy. We believe that the same is true for economic macro-systems.

How do such dissipative structures appear? They form as a result of random fluctuations which force the system out of a state of thermodynamic balance, provided there is an influx of energy and matter into the system. (What happened in the U.S. economy after September 11th, 2001 was not characteristic of a state with minimal entropy). According to the second principle of thermodynamics, closed systems are at risk of "thermal death". In thermodynamic terms, the U.S. government tries to do everything it can to make the U.S. economy open. Possible evidence of this, at least, is the war being fought for energy to inject into the U.S. economy - the energy in this case being oil from Iraq. We will touch on the possible analogy of the "thermal death" of the U.S. or the world economy in the closing part of the article.

Like it or not, our internal energy reserves (oil, gas, uranium etc.) will run out at some point. It is impossible to keep injecting energy into the economy infinitely.

Sadly, the limits of this article do not permit us to set out a "general course in econothermodynamics", so we will attempt to apply our proposed econothermodynamic approach to describe just one of the structural units of economics - a business company. How universal are the processes we are discussing?

## ECONOTHERMODYNAMIC PARAMETERS

The present publication is aimed at a broad spectrum of readers, so while we are using thermodynamic principles to describe commercial companies, we will use everyday language for our explanations and may make somewhat free with physics terminology (*pace* physicists!) And to serve our purpose, we may use not only thermodynamics, but also the general laws of physics, since they are valid for natural phenomena. Thermodynamics is convenient in that it allows us to characterize the development of systems on a macro-level without going into fine detail. Thermodynamics introduces the concept of an ideal gas. Conventional thermodynamics often analyzes closed systems, where the amount of matter is constant and no energy is exchanged with the environment. Let's assume that the macroeconomic system we are looking at is an open system where there is an "exchange of energy and matter with the environment" (here we have in mind actual energy - oil, gas, electric and atomic energy, and also natural resources used in the economy). According to Prigozhin, such a system, developing according to natural principles, will tend towards a state with minimal entropy growth.

Here a vital question arises: how do we correctly introduce the concepts of entropy and temperature and other thermodynamic parameters for our macroeconomic system? V.P. Maslov [11] introduces the concepts of entropy, temperature, free energy and the Hamiltonian into problems to do with the theory of probability for identical objects. This allows author to apply the latest methods of quantum statistics to financial problems.

We believe that it is important to establish the scale of the systems concerned. For example, what is one mole for an ideal macro-economy? Maybe the ideal macro-economy does not exist, just as there is no ideal gas? Unfortunately, this is probably the case. We will attempt to answer this question in further publications. Let us assume that we are looking at an econothermodynamic macro-system (for example, the macro-economy of a country). We will attempt to describe the econothermodynamic characteristics of the system using normal thermodynamic parameters. For example, when we are talking about the internal energy of the system $U$, we can introduce the concept of absolute temperature $T$ as a econothermodynamic parameter to describe the state of thermodynamic balance of our system, related to its internal energy $U$ (econothermodynamic function of state of the system). In this way, it can be assumed that for a system in the state of balance, this particular thermodynamic parameter is identical in all parts of the system. For example, decreasing of the internal energy $\Delta U$ is equal to the work $A$ preformed out during reversible process by an adiabatically isolated system on internal surroundings (for example, by the "vacuumized" economy of North Korea on economics in touch). An adiabatically isolated system may be not only a system in a vacuum (where there is no exchange of heat with the environment), but also a system where all changes take place very quickly (where the system simply does not have time to exchange heat with its environment, within the normal time-limits of the process, or where the exchange is negligible). The internal energy $U$ of such a system includes the internal energy of each object in the system and the energy arising from the interaction between individual objects in the system. In a practical sense, internal energy $U$ may be a way of describing an economy's potential for future expanding and development, while temperature, following the same analogy, may be a measure of the level of an economy's work activity or its general health. Please, note that in accord with upper definitions the temperature $T$ – a parameter of the system in a balanced state – a measure of economic health of system when it transforms from one balanced state to another. The balanced state of the system with different temperature in different regions will be the situation with unhealthy economics. As example, we can consider Chinese economics with quickly developing large cities and poor agricultural regions. Of course certain temperature different can exist because even for an absolutely healthy person the temperature of axillary region and nose are different. Let us assume that there is a constant *κ (analogous to the Boltzmann constant)* which links these two values and provides a link between the micro- and macro-levels of a company. The heating up or cooling down of a macro-economy may occur as a result of a whole range of internal and external micro-processes.

For example, reducing the bank rate is a method widely used by the governments of some countries when attempting to increase the temperature of the country's macro-economy. The question is, how is a given influence connected on a macro-level to changes in other econothermodynamic parameters.

Pressure, **P,** in econothermodynamics can generally be defined as a scalar value characterizing the level of stress or loading in a macro-economy. As in conventional thermodynamics, pressure in a macro-economy can be defined through the infinitesimal work *dA* performed by the macroeconomic system during a quasi-static change (increase or decrease) in its volume *dV*: *dA=PdV*. The work $A_1$ performed by the macro-economy of a particular country on the external macro-economy and the work $A_2$ performed by the external macroeconomic system on the macro-economy of that country will be opposite in sign $A_1 = -A_2$. Certainly, average value of pressure also characterizes the mean normal forces with which the macroeconomic system affects the unit of area surrounding its business space. In an everyday sense, external pressure *P* can characterize the force of the influence of a one macro-economy or its elements in the certain direction on other structural units of the world's economic area surrounding the macro-economy from this side. The internal pressure also indicates the kinetic energy of individual structural units of the macroeconomic system. It would seem that atmospheric pressure and temperature corresponds to the situation where the system has normal health and is working stably and harmoniously, and the internal and external influences are balanced (in other words, the macro-system is "enjoying" the state of comfortable internal and external economic conditions).

It is also possible to introduce an analogy to, for example, the magnetic field *H*, **corresponding to the competition** between external and internal (trading) interactions in the economy. This competition can lead to the regulation of vectors of individual structural units making up the system (for example, if we take the economy of Europe as a whole, we can talk about regulating the direction of development vectors of the economies of individual member countries of the European Union) or (at certain temperatures) it can lead to chaos within the system (even though the system itself continues to exist as a single macro-economic system).

Volume, **V,** characterizes the spatial size of the system, for example, the area occupied by the macro-economy of Europe.

We would like to note once more that the limits of this article do not permit us to set out a whole "general course in econothermodynamics", and so the definitions we have given of temperature, pressure and volume are rather qualitative. Our aim in this work is to evaluate the possibility of applying a thermodynamic approach to macroeconomic systems. If this approach is taken up by the academic community, we will try in future publications to describe more precisely the connection between these thermodynamic parameters and specific macroeconomic indices, or indices of the work of an individual country's macro-economy.

*Phase transitions*

We do not claim (and it would be somewhat strange if we did) that a business community can only develop according to the laws of nature. Macroeconomic systems (like the real activity of individual large companies) are very complex, and since it is impossible to acquire completely comprehensive information about them, it is not possible to postulate formulas for their development. But we suggest taking these natural laws into account in the macro-economy. The possibility of their application should by all means be checked and tested scientifically, but we cannot afford to ignore them. If a complex open system, such as our macro-economy, is to survive and develop successfully, it must learn to endure "catastrophes" and navigate safely past bifurcation points, near which it will be in an unbalanced, unstable state. Or, in physical terms, the system must endure phase transitions, when it moves from one phase to another because of changes in the parameters relating to its state as a whole (order parameters) and describing its thermodynamic balance. (Such a transition might be, for example, the re-orientation of an economy or its complete reconstruction after destruction by a crisis or by war). Phase transitions are widespread in nature. For example, some first-order phase transitions are physical processes such as evaporation, condensation, melting, sublimation, etc. In other words, during first-order phase transitions some physical properties (density, concentration) change abruptly and a certain amount of (latent) heat is released or absorbed. During second-order phase transitions, no release or absorption of heat is registered, and density and concentration change constantly, not abruptly, but an extra physical variable (or order parameter) must be defined to describe the state of the system. This order parameter is equal to zero on one side of the critical or transition point, but on the other side it gradually increases as it moves away from the critical point. Leo Landau (1937) suggested that all second-order phase transitions could be interpreted as transition points at which a system's order of symmetry changes. In physics, these would be phase transitions where processes occur giving rise to superconductivity in metals and alloys, superfluidity in helium, spontaneous polarization in ferroelectrics and magnetic moments in magnetic materials.

A macro-economy can undergo analogous phase transitions when, in the course of its development, it moves from one balanced state to another; i.e., from one type of structure to another, which may even be an ideal type. Of course, it should be noted that different types of event constitute phase transitions for different systems. For some systems a phase transition may be a change in the macroeconomic structure, for example a switch from a raw-material economy to a high-technology economy (a structural phase transition); for other systems it may be the introduction of a reduced bank rate (an increase of temperature and pressure in a system with limited volume) which keeps the system from transition into an unregulated state (the relocation of the system to a phase space farther away from a phase transition). In physics, such qualitative changes may be structural, magnetic or electronic phase transitions, and these can also occur simultaneously. Similarly in economics, phase transitions in different sub-systems of a macro-economy can coincide (for example, the complete collapse of an economic system caused by a revolution with a phase transition within the social sub-system).

It is obvious that during the development of a macro-economy it is very important to avoid first-order transitions - abrupt, revolutionary changes (as well as actual revolutions). The same can be said about large business companies. Successful companies are the ones which have developed smoothly, their employees sometimes not even fully aware of the nature of the changes taking place. This observation is illustrated by a five-year research study into the activity of successful business companies led by Jim Collins [12]. Firstly, one of the main conclusions the economists came to is that practically any organization can significantly improve the results of its activity and possibly even become great.  However, Collins concludes that "Those who begin revolutionary changes, major restructuring and transformation projects, will definitely fail to achieve outstanding results. Whatever the final goals are, the transition from good to great cannot be achieved in one leap" [1]. This means that bosses undertaking a first-order phase transition are running a grave risk, because they are trying to take the system into a new thermodynamic state, in which it has never been before.  But that is not all. First-order phase transitions usually take place very quickly, and for this reason even physical systems often do not have time to stabilize. They either overshoot the stable state, or do not reach it, getting "stuck" indefinitely a metastable state, till the next shake-up. The main problem is that it is impossible to determine the energy needed for transition into the new stable state without knowing the exact "energy spectrum" of the new state and the exact location of its "local energy minimums". As Collins would say, a company which announces huge restructuring and reorganization is, in fact, sentencing itself to failure. So the important conclusion is this: "life" prefers second-order phase transitions (smooth transitions without shake-ups, but which are as fast as possible under the given circumstances). Any driver knows that the worse the lighting on the road is, the more slowly you should travel, especially at sharp bends… but go forward you must.

The success of a macro-economy can also be considered (at least partially) as a consequence of its basic, inbuilt processes and its fundamental dynamics. What are meant here are the processes which are an integral part of the system itself, not the results of a single big idea or the actions of one great, omniscient leader making important decisions. Actually, the history of many countries is rich in reverse examples, where charismatic leaders with unrestricted power, or power-seeking revolutionaries, have led the economy to total chaos while trying to accomplish the usual sort of revolutionary changes (first-order phase transitions).

So, despite the fact that in physics the thermodynamic approach cannot explain micro-level processes, it can be successfully applied to characterize phase transitions. This applies to relatively simple three-dimensional objects consisting of separate atoms and molecules, the size and shape of which, as they approach to a phase transition, (for example, a three-dimensional magnetic, but possibly even a company) do not affect its type and the point at which it occurs. It is also true for quite complex structures whose shape does change during a phase transition - for example, for an alanine polypeptide [13].

*Development of macro-economies, order and stability*

How should the activity of a macro-economy best be organized in order for it not only to survive without crises, but also to become highly successful? How should an open, developing, thermodynamic system of this type behave?

A leader of national economics, trying to establish an ideal structure for his country's macro-economy "brick by brick" may be wrong in thinking that such a system will be the most long-lasting and stable. He may indeed end up with a structure which, from a physics point of view, is able to withstand external influences (or at least the influence of pressure **$P$** and temperature **$T$**). Like a crystal or like matter with a crystalline structure, his system will require a much higher pressure and temperature to destroy (crash or melt) it, but a structure like this is not "alive". On the one hand, such a structure is very convenient. Every structural element knows its place and its role. Its interactions with other structural elements are regulated by laws and instructions. But in a real-life situation, a regulated system of this sort is less flexible and mobile. It is more difficult for it to adapt to changing conditions of life, and some kinds of influence will make it highly unstable (requiring the rewriting of a lot of laws). Its analogue in physics is a crystalline structure which is unstable in relation to some kinds of deformation, e.g. shear deformation. Regardless of how hard it is (perhaps as hard as a diamond), it has, unfortunately, zero flexibility and is often adapted withstand only two kinds of influence – high pressure and temperature.

Amorphous bodies have more stable structures than crystals. In contrast to crystals, which exhibit long-range order, amorphous and vitriform bodies exhibit only short-range order in the arrangement of their atoms. Such materials preserve the high degree of symmetry inherent in liquids, even though their atoms are in fixed positions. They possess local bonds and local anisotropy (e.g. magnetic). The peculiarities of their structure define the specific physical properties of amorphous materials. For example, although the density of amorphous metals is lower than that of crystalline ones, because they are less densely packed, their strength is 5-10 times higher. This effect is attributed to a lack of conventional structural defects in the amorphous state. Amorphous bodies are more stable than crystals: stretched in any direction, their bonds do not break. In economic terms, it means that in a macro-economy organized in an "amorphous" way, with local short-range order, unlike with a crystal, it is impossible to pick a single angle at which to "hammer" it in order to break it. It would seem that to achieve such an "amorphous" macro-economy, there should be stable energy states forming a stable microstructure with a defined short-range order. At the same time, some local disorder can be allowed in the local environment, to which the state in question must be firmly connected. Only at high temperatures can this lead to re-crystallization (i.e, to the crystalline structure mentioned above). Local disorder can be, for example, the absence of over-strict government regulation of economic processes for each element of a macroeconomic system (i.e. a liberal economy). In the case of large companies, it can be an absence of the over-detailed instructions which some bosses try to use instead of trusting the 'common-sense' or self-discipline of their employees.

So what structure is the most viable? There is no need to reinvent the wheel – nature has already created the structure we need. It is the DNA molecule - the carrier of the code governing the chemistry of all living things.

### *DNA*

The deoxyribonucleic acid (DNA) molecule is a right-handed spiral consisting of two polymer chains connected with hydrogen bonds. Each polymer chain consists of structural units called nucleotide elements connected to each other with phosphodiester bonds. DNA comprises four nucleotides: adenine (A), cytosine (C), guanine (G) and thymine (T). The nucleotides of two polymer chains form the DNA molecule based on a principle called complimentary pairing: A is connected to T (with two bonds), G is connected to C (with three bonds).

This is the most viable structure to have been passed down by evolution, the most adapted to external influences such as temperature. DNA carries out its own sort of "information recording", just like a DVD-disc or a flash drive with files. If a dog's tail is cut off, it doesn't mean that the tail gene from the DNA disappears with it. So the puppies of that dog will still have tails. Similarly, after the 9/11 disaster, a structural element of the U.S. economy (aviation) was delivered a serious blow. But the information "kept in the DNA" of the U.S. macro-economy at the genome level allowed the aviation sector to be rebuilt.

It is precisely in dealing with the main "disordering" factor - thermal energy $\kappa T$ - that DNA exhibits its stability, protecting and preserving information about the living organism. Within the DNA molecule can be found elements of the icosahedral symmetry characteristic of living systems and of the golden ratio (in cross-sectional view). DNA lacks the sort of defined phase transition that takes place, for example, during the melting of a crystal such as ice. DNA melts step by step when heated, but its individual sections are melting at different temperatures. The development of a DNA molecule and the preservation of its integrity are very complex processes. Like everything "physical" in the universe, it obeys the principles of thermodynamics. And the hierarchy of its disorder growth complies with the second principle of thermodynamics.

Basically, the development of any living system involves the constant struggle against entropy (the disordering effect $\kappa T$). According to Prigozhin's theorem of unbalanced thermodynamics, in the presence of external links which prevent it from reaching a state of balance, a thermodynamic system will achieve a stable state when it produces a minimum amount of entropy. Thus life, in such (dissipative) systems, introduces the minimum amount of chaos into the universe when it is in a balanced state.

*Application to large business companies*

If we look at the macro-economy of a large business as being a system like this, then, in one way or another, the activity of its structural units will lead to an increase in entropy. In order to maintain stability, energy must be expended to reduce entropy and bring about order. Bearing this in mind, a company boss needs to make a correct assessment of his options and expend less energy on combating entropy (bringing about order), and more on developing new products and services and bringing them to market. On the other hand, order does have to be established. A company without order cannot create new products. A boss who expends all his energy on reducing entropy (dealing with employees' training, organization and everyday problems) will prove an ineffectual business manager. Yet someone must do this, otherwise the company will collapse (fall into a state of maximum entropy). In our view, it is obvious that a business structure cannot be built like a crystal, because this leads to a "dead" system, in which sooner or later, when external variables alter, a phase transition will take place. A system needs to be "alive" in order to adapt to changing external conditions, and it must be sufficiently flexible to withstand external influences.

There must be well thought-out internal connections. Take the example of DNA, where such connections are well-organized and able to react even to increased temperatures, by changing their structure slowly and over a long period of time. The connections within a company must also be very strong yet mobile. If the company covers a variety of business lines, these lines must be both sufficiently independent and at the same time strongly interconnected, like the multiple cross-connecting hydrogen bonds in DNA (see also the final section of this article, about the correlation between kinetic and potential energy within a company).

Let us return to Szilard's conclusion about the circumstances in which a rational being is able to break the second principle of thermodynamics. According to the author, he "must be accurately informed about the state of the thermodynamic system". Reasoning from this, we may assume that, given a well organized body of information (with an efficiently updated data base and swift transfer of information from top to bottom and vice versa) it is possible to "diagnose fluctuations" (steps or tendencies) in the wrong direction and to give them a "uni-directional alignment", thus breaking the second law of thermodynamics for closed systems. And it follows from this that it is possible to achieve a reduction in entropy, in effect turning the closed system into an open one (where an exchange of energy and matter, including information, takes place with the external world). Since human beings are basically open systems, there can be no doubt that business communities should also be constructed on an open model. However, one must note here that (just as in the case of "Maxwell's demon") a certain amount of energy is expended on the identification and gathering of information, as well as on managing it, and this expenditure cannot be compensated for in a closed system. In other words, it is impossible to accomplish the above (to break the second principle of thermodynamics) in a closed system.

But what if this exchange of information, energy and matter is disrupted or malfunctions? What threat will this pose to the company? Why is it a bad thing if a company acquires the characteristics of a closed system? Because if it does, "thermal death", which is characteristic for closed thermodynamic systems, will be inevitable. The collapse of a large 'closed' corporation can only be prevented if the corporation learns to recognize "fluctuations" (steps in the wrong direction) and to accurately determine its thermodynamic state even without comparison with the outside world (otherwise managers will write unrealistically sugared and glossed-up reports to their superiors, bearing little relation to the real state of affairs). While a slightly ill person can take his own blood pressure and temperature and diagnose his illness, when he is in a grave condition, even a doctor sometimes cannot diagnose and treat him appropriately. We believe that it is impossible to evaluate the state of a company without comparing it to the outside world: the market is the external object that puts the state of the company into perspective. It is the existence of this external, independent "evaluator" which allows the development of the liberal market economy: this is "the invisible hand" (A. Smith) that guides human economic activities. The exclusion of this factor from the economy, we believe, turns it into a closed system, which in the end will lead to its collapse. An evaluation of other interpretations (the revelation of divine providence or the consequences of the probabilistic nature of the world) does not lie within the scope of the present publication.

*Viability of small business companies*

In physics, the size of a system can influence its properties. At some critical sizes, for instance, its surface or the restructuring of its energy spectrum can play a decisive role. For example, when the size of a system changes from one bulk substance to a group nano-objects, the whole structure of the material (crystalline, electronic or magnetic) is altered. Furthermore, non-magnetic matter can turn into magnetic matter. The same thing happens during the transition from a small business to a large one: both the structure of the system and the management methods change.

Physics, like other sciences, actively explores the nano-world (1 nm = $10^{-9}$ m). Although the term "nanotechnology" is currently enjoying unprecedented popularity, even in historic times nano-particles of gold were added to the stained glass of catholic churches in Cologne. Modern scientists have established that using nano-objects in technology and medicine is usually much more effective than using bulk materials. It would appear that within large businesses, small "nano-particles" or "nano-clusters" may possess much greater efficiency. And even a monopoly, if it consists of small, mobile and adaptable nano-clusters, will function with a more organized and flexible approach to management. Our estimates indicate that, from technological, psychological and structural points of view, the optimal size of such "nano-companies", managed on two levels, is approximately 10-100 employees. This corresponds, with amazing exactitude, to the number of atoms in a spherical particle of about 1 nm in size. The surprising chemical activity of such particles is influenced by the fact that the percentage of surface atoms is about 86%. In small companies, the vast majority of front-line employees are very active (they cannot hide behind their colleagues). This fact largely explains the higher labor efficiency in such companies. Naturally, large and small companies have differing functions within the macroeconomic process, and it is hardly possible to organize mass production with only 100 employees. However, it is now established that bulk materials with a nano-structure usually have markedly more advanced properties than conventional bulk alloys and compounds. (Nano-structurized objects are materials with the nanometric size of constituent crystallites (1 - 100 nm). Due to the small size of crystallites (size of a crystallite in a conventional material is > 1 μm), a large part of atoms in such materials is located in intercrystallic (intergranular) barriers, which defines their unique properties.)

Bearing this in mind, perhaps giving large companies a nano-structure could lead to positive results? Which is more economically efficient - to create a number of small companies or to give large ones a nano-structure? We believe that this issue remains open, although the macro-economies of developed countries favor the former.

In future publications, we will attempt to demonstrate mathematically the possibility of applying thermodynamic and other physical laws to the macro- and micro-level of the economy. In conclusion, we would like to say a few words about the concept of "resonance frequency". In physics, this concept can be applied to basically any system - from atoms, molecules and radio circuits to bridges, large buildings and complex non-linear systems. In work [14] the concept of resonance is applied to the educational process. Everyone knows the classic example from school textbooks: a squadron marching over a bridge is asked not to march in step, in order to prevent the effect of resonance, with a frequency of about 1 Hz, on the bridge supports. Is it possible to apply the concepts of resonant or natural frequency to the macro-economy or to business structures? We believe so. In these cases, resonance occurs when a macroeconomic system or the personnel of a company are especially sensitive to the influence of an external force when this influence occurs at a certain frequency. If the natural frequency of the system and the frequency of the impinging influence coincide, this can lead to a sharp rise in the amplitude of vibration in the system's individual structural units. If we now look at systems with changing properties (rapidly growing companies or economies), we can assume that the value of the resonance frequency can change. For example, in companies with a small staff or a low turnover (organizations with a small effective mass), the resonance frequency is usually comparatively high. It is determined not only by the number of employees, but also by their desire to work, their intelligence and self-discipline, and by a number of other factors, including geographical ones. The expenditure of energy needed to disseminate information throughout the organization and to organize employees' activity is small in this case. It means that irreversible losses through dissipation and attenuation are not significant. And if there is a "snag" that requires some effort on the part of the boss (effort spent on reducing entropy instead of on constructive activities), it is easy to resolve. As the effective mass grows, the significance of resonance frequency starts to decrease. We also know that it is quite hard to influence large objects using a low frequency (0.1 Hz or even much less), because this demands a fairly high expenditure of energy. It is precisely this fact that can lead to a sharp increase in administrative personnel and/or even to the collapse a huge corporation after the departure of a powerful or charismatic leader if he constructed "brick by brick" company instead "alive" one. For this reason, it is systems with a small staff and an optimal resonance frequency (from the point of view of working capacity and viability in the prevailing conditions) that are the most manageable and long-lived. Of course, they do also have a number of disadvantages when it comes to surviving in difficult conditions, e.g. insufficient resources and highly-paid personnel. One can assume that it is more energy-efficient to build corporations with "nanostructures" based on a cluster system [15]. It is much more profitable for a large corporation to grow a grove of trees than to lavish care on one gigantic tree that can be destroyed by a single bolt of lightning. Organization based on a cluster system allows a much speedier response, for example, to changes in market conditions or the introduction of new products.

Naturally, the question may arise as to how to manage structural units in a cluster system if they have different resonance frequencies, how to organize links and cooperation between them, and how they can support each other. There will also be the question of how to align the vectors of the clusters in the correct direction. What means of influence is it permissible to apply in order to regulate these vectors? Our next publication will be devoted to the general question of how to coordinate all the structural elements of a cluster system so that they work harmoniously together.

Let us return to those companies whose managers are constantly faced with the problem of small teams breaking away (or trying to do so) in order to form their own businesses. Let's review this issue from a thermodynamic point of view. We must understand the difference between the personnel of a company and the man in the street. The former possess some degree of interaction and team structure. There are corporate rules in place, each employee has a superior, each has official duties, has been taught working practices, given a workplace, and allocated to an organized department. The company, then, corresponds to matter in a condensed state where rigidity is not zero. If the kinetic energy $E=mV^2/2$ of a particular business department ($m$ is effective mass of this department in the company and $V$ is the mean velocity) is significantly higher than the potential energy $U$ (here we must understand potential connections as being the whole complex of departmental interactions, from informational to technological), then separation is inevitable. Putting the brake on (reducing speed $V$ with quadratic dependence) is the method widely used by bureaucrats at different levels, but is it helpful to the company as a whole? A car cannot be controlled without brakes, although it can move without them, and very fast too (or can but only one time). We know that inertia laws are not only applicable to mechanics. Would it perhaps be preferable to concentrate, not on slowing it down, but rather on reducing the relative effective mass *m* of the department with respect to other departments (so that a separation will not be critical) or on increasing the strength of potential connections (by more serious involvement or attachment of the department to the activity of the company as a whole)? That is exactly how matter behaves in a condensed state, by taking on solid form (where the objects of which it is composed have low kinetic energy *E<U*), liquid form (*E~U*) or gaseous form (*E>U*). As mentioned above, such a "liquid crystal" phase is one of the possible options for a company structure. Another (and probably better) possibility is where the departments and sectors are well connected, like DNA with its multiple interconnecting hydrogen bonds.

**CONCLUSION (ADVICE ON THERMODYNAMICS FOR A COMPANY DIRECTOR)**

We do not have economic degrees, so we're not in a position to give advice to the directors of companies and/or leaders of macroeconomic processes. However, we've come to the following conclusions regarding the activity of individual companies from a thermodynamic point of view.

Firstly, when managing a company, **don't waste all your energy on combating entropy (disorder)**. Concentrate also on creation and developing of new products and services. Find the golden mean and correctly distribute your energy between combating disorder and being creative. The director should organize the activity of the company in such way that the company structure itself will combat entropy growth (in the ideal scenario) or delegate specific people to this task, or else ensure that entropy growth will be inhibited of its own accord (the latter can also be a bad sign, if the entropy has reached its maximum).

Establish an exchange of information within the company (efficient updating of the database, quick transfer of information from the top down and vice versa) and a similar exchange of news and communication with the outside world. In this way, according to Szilard, it will be possible to "diagnose fluctuations" (steps or movement) in the wrong direction and give them a "unidirectional alignment". The second principle of thermodynamics for closed systems will thus be broken and, having achieved a reduction in entropy, it will be possible in effect to turn the company into an open system.

Even a huge business company must be an open system to be sure to avoid "thermal death" (which, according to the second principle of thermodynamics, is the unavoidable fate of a closed system). A human being is a typical example of an open system. A company must constantly be **open** to the exchange of energy and matter with the environment.

Secondly, **avoid "saturation"**. **A vegetable existence takes more energy.** There is no such thing as a "finishing line" in a great business company! Sooner or later all physical systems become saturated, with a high probability of a first-order phase transition. Therefore, it is those companies that develop smoothly and steadily that are successful. "The transition from good to great takes enormous effort, but this effort allows potential to be accumulated, which at some point starts to give back more energy than was invested in the beginning", says Collins.

Carefully think through the structure of your company in the context of the influence on it of all thermodynamic factors, and develop a **system of internal connections** between individual structural units, departments and employees. Our advice is: arm yourself with the sort of system which contributes best to the company's viability (like the DNA molecule).

**"Life" prefers second-order phase transitions** (smooth reversible transitions without shake-ups, but as fast as possible under given circumstances). Do not make revolutionary changes to the parameters defining your company (e.g. its structure).

From technological, psychological and structural points of view, the optimum size of most effective small companies, with two levels of management, is approximately 10-100 employees, which amazingly coincides with the number of atoms in a spherical particle about 1 nm ($\sim 10^2$) in size. The leaders of large and medium-sized business companies should probably consider converting their business to a nano-structure.

How can you find the golden mean between reducing the effective mass of departments (or other structural business units) and increasing the rigidity of potential connections? It is up to you to decide in each specific case how best to retain the maximum kinetic energy of your employees and departments without the company falling to pieces. You will probably be unable to avoid small slowdowns (they can happen due to increasing of potential connections), but they can actually be most effective.

In conclusion, a few words about the "thermal death" paradox. By "thermal death", we mean precisely the phenomenon described by Professor Meadows' group: a sharp decrease in the volume of industrial production, demographic decline (as a consequence of insufficient food production and a decrease in life expectancy), and an initial increase, followed by a decrease in the level of environmental pollution, due to the cessation of production, etc. Is this a reality, or is the thermodynamic approach not applicable? There are no sure facts supporting the idea of the Earth as an open system. By the way, the paradox of the "two twins, which follows from Einstein's general relativity theory formulated about a century ago, is still discussed in scientific circles. Obviously, there is still enough energy and matter (oil, gas, uranium and ore) on Earth, but these resources are limited. As an example, authors of work [16] conclude, the world's oil production is entering the stage of irreversible decline peaking in 2006-2010  What will happen then?

We believe that nuclear energy, steam/gas and coal power stations not only will not save the situation from the point of view of entropy, but will even aggravate it. What then? Hydroelectric plants, hydrogen energy, modern polymer solar batteries, hypereffective windmills, magnetic heat pumps? We must develop these types of energy very rapidly, but we must also understand that even then the larger orders of magnitude are required. Maybe the chemists will synthesize new materials and physicists will finally solve the problem of controlled nuclear fusion. Billions of euros have already been spent to solve this problem and dozens of billions will be spent in the next few decades (on the scale of world economics, it is not a very large sum), although it is not even theoretically clear yet whether thermonuclear fusion complies with the second principle of thermodynamics or not. Will this solve the problem of limited resources of ore and other chemical compounds whose treatment processes are irreversible? Or will the humanity be forced to replace elution at the mining and concentration plants with gathering single molecules of iron, nickel and cobalt pulverized over the surface of Earth after multiple usage?

Is it possible that solving the problem of controlled nuclear fusion is the only possibility to receive an inexhaustible source of energy (quantitative measure of movement and interaction between all kinds of matter)? We are sure this is not the case. Firstly, we doubt that it is theoretically possible to contain stable plasma at the temperature of about 150 million K in Earth conditions for a sufficient period of time. Furthermore, it is hard to imagine the possible consequences of a smallest malfunction. And where can we find the materials necessary to sustain already uncontrolled nuclear fusion e.g. in case of malfunctions during the cooling of superconduction solenoids, or errors in configuration of the magnetic field necessary to contain the fusion? Secondly, it is not clear why do we need to create a small, controlled, "tamed Sun" (plasma sphere) on Earth? The fact is that 7 of 21 light nucleus reactions of interest are basically the schemes of fusion bomb reactions (including the $d+t \rightarrow {}^4He+n$ reaction). The Sun mass constitutes 99.87% of the mass of Solar system. That means that the Earth mass (333000 times less than the Sun mass) is negligible. Maybe there is no reason to create a "controlled fusion bomb" in our backyard, and it would make more sense just to increase the usage efficiency of the inexhaustible energy source that has been existing for billions of years and has radiation power of $4 \times 10^{26}$ W? About $10^{18}$ W reaches the Earth surface daily. But most of this energy is transformed into heat, reflected off the Earth surface and returned into space, and solar batteries cannot solve the problem in full. It is necessary to develop complex approaches to the development of highly efficient solar energy technology, including heat pumps and most importantly, magnetic heat pumps, which in accord with report published by the Prof. P. Egolf group are most efficient [17]. With financial backing comparable to that provided for the development of nuclear fuion power sustem ($10^9$-$10^{10}$ Eur), the mentioned approach is bound to bring desired results.

In future, man will begin to actively colonize other planets of the Solar system in search of additional material resources (although some of the closer planets are probably already in the state of "thermal death"), therefore we should begin to consider the physical and technical aspects of energy accumulation and/or energy transfer over large distances. Then, we believe, we would have an opportunity to avoid "thermal death" of the world economy. It can be done by developing new physical and chemical methods of highly effective transformation of heat energy received by Earth from the Sun (e.g. into electric energy) and/or developing highly effective methods of accumulation of vast amounts energy and its transfer over large distances. Of course, the professionals can laugh and ask us how are we proposing to materialize these ideas. Being specialists only in the thermodynamics and physics of magnetic phenomena, we are prepared to give specific recommendations only regarding the first part of the issue, the heat energy transformation. But would it be reasonable to decide the fate of the world depending on whether or not the scientists will be able to keep plasma under control? And then, how long will it last? It is not right to decide the fate of the world based on the success or failure of a single scientific project.

Therefore, we must heavily reduce the amount of irreversible processes on Earth as soon as possible in order to combat the increase of entropy in the world economy. But since the factories are not being shut down, we will have to transport the matter, which would take energy in amounts much higher than are available today. We would need to be talking about generating hundreds of terawatts (possibly even accumulated outside of Earth) and using this energy to transport matter (iron, titanium, aluminium) to Earth (possibly even $^3$He to carry out environmentally safe nuclear fusion reactions)

The present publication is a first attempt to draw parallels between thermodynamics, economics and business structuring. The authors would be happy if not only physicists, but also business leaders and macro-economics specialists would join the present discussion.

The authors would like to thank Professor Dr. B.A. Strukov, Cand.Sc. Y.I. Spichkin and student V. Zverev for instructive discussions regarding the possibility of applying a thermodynamic approach to macro-economics.

authors do not like so much to use the term "nano" and "nano-structuring" of business companies. It can be considered as an attempt to use popular terms to get more publicity.

16. Nigmatullin R.I. and Nigmatullin B.I. [http://www.proatom.ru/modules.php?name=News&file=article&sid=336](http://www.proatom.ru/modules.php?name=News&file=article&sid=336).
17. http://www.bfe.admin.ch/dokumentation/energieforschung/index.html?lang=en&publication=9131.